\documentclass{article}
\usepackage[utf8]{inputenc}
\usepackage{url}
\usepackage{graphicx}
\usepackage{multirow}
\usepackage{xspace} 
\usepackage{framed} 
\usepackage{mdframed}
\usepackage{multirow}
\usepackage{makecell}
\usepackage{caption}
\usepackage{subfig}
\usepackage{comment}
\usepackage{listings}
\usepackage{xcolor}
\usepackage{authblk}

\title{Accurate Measurement of Application-level Energy Consumption for Energy-Aware Large-Scale Simulations}

\author[1]{Osman Seckin Simsek}
\author[2]{Jean-Guillaume Piccinali}
\author[1]{Florina M. Ciorba}

\affil[1]{University of Basel}
\affil[2]{Swiss National Supercomputing Center}

\begin{document}

\maketitle

\begin{abstract}
    Sustainability in high performance computing (HPC) is a major challenge not only for HPC centers and their users, but also for society as the climate goals become stricter. 
    A lot of effort went into reducing the energy consumption of systems in general. 
    Even though certain efforts to optimize the energy-efficiency of HPC workloads exist, most such efforts propose solutions targeting CPUs. 
    As HPC systems shift more and more to GPU-centric architectures, simulation codes increasingly adopt GPU-programming models. 
    This leads to an urgent need to increase the energy-efficiency of GPU-enabled codes.
    However, studies for reducing the energy consumption of large-scale simulations executing on CPUs and GPUs have received insufficient attention.
    In this work, we enable accurate power and energy measurements using an open-source toolkit across a range of CPU+GPU node architectures. 
    We use this approach in SPH-EXA, an open-source GPU-centric astrophysical and cosmological simulation framework. 
    We show that with simple code instrumentation, users can accurately measure power and energy related data about their application, beyond data provided by HPC systems alone. 
    The accurate power and energy data provide significant insight to users for conducting energy-aware computational experiments and future energy-aware code development. 
\end{abstract}

\section{Introduction}
\label{sec:intro}

Energy efficiency is a major concern in high performance computing (HPC), as it can prevent significant operational costs and environmental impacts.
Reducing energy consumption in HPC systems requires the coordinated efforts of hardware vendors, system designers, and HPC users. 
Hardware vendors have made significant progress in improving the computational efficiency of HPC systems, leading to a 15-fold increase in efficiency in the Green500 list over the past decade~\cite{Green500}. 
HPC centers also took steps to reduce energy consumption, e.g., switching to renewable energy resources. 
However, more can be done to improve energy efficiency in HPC systems. 
HPC users can also contribute by optimizing their simulations to run more efficiently in terms of performance and energy.

Measuring the energy of a simulation requires power and time measurements.
Although external power meters are the most accurate method to monitor the power consumption of computing systems~\cite{romein2018powersensor}, power meters are generally not integrated in production systems.
Instead, the systems provide the power and energy measurements through embedded sensors using IPMI or ACPI interfaces.
However, the power consumption data is only available to HPC system administrators and not to system users.

In HPC systems, resource and job management systems, such as Slurm, can be configured to provide information about the energy consumption of submitted jobs. 
Specifically for Slurm, this information can be enabled using the Slurm Energy Accounting Plugin (AcctGatherEnergyType). 
Depending on the system, the energy plugin back-end can be \textit{ipmi}, \textit{pm\_counters} or \textit{rapl}. The \textit{ipmi} uses the baseboard management controller~(BMC) and \textit{pm\_counters} is an HPE/Cray-specific BMC, while \textit{rapl} uses the hardware sensors to read the power consumption of the CPUs.

The data collected by the Slurm can be accessed by the user with the \textit{sacct}
command at the end of a job's execution. 
While energy consumption data provided as part of a job report is useful for users, it does not provide specific information about which parts of the job consumed most energy, depriving the user of the opportunity to reduce their job's energy consumption.
When information is available about devices that consume more energy, or as a functional breakdown, users can then employ various techniques to reduce their application's energy consumption. 

In this work, we show the benefits of enabling accurate measurement of energy consumption and reporting this information for each compute device and simulation function to the user. 
We showcase our approach on a real-world use case from computational astrophysics, namely the SPH-EXA simulation framework~\cite{cavelan2020sphexa}, executing on three different CPU+GPU node architectures. 
We integrate an open source power measurement toolkit \cite{2022pmt} (PMT) into SPH-EXA, to gather energy consumption information, and create reports that users can analyze for possible energy-reducing code development and computational experiments.

\section{Methodology}
\label{sec:methodology}
\textbf{Energy measurement.} PMT~\cite{2022pmt} is an open-source library that provides a common interface to measure power consumption of various devices.
Using the power measurements taken from the vendor provided APIs, PMT calculates the energy consumption of the instrumented code and reports the measurements to the user. 
The advantage of PMT compared to previous efforts for power measurements, e.g., LIKWID~\cite{treibig2010likwid}, is that it provides an interface to a comprehensive set of back-ends which reduces the frequent code changes within the application code base.
PMT also supports the HPE/Cray-specific back-end, thus providing easy and accurate measurements for compute nodes entirely built by HPE/Cray.

The HPE/Cray PMT back-end uses the HPE/Cray-provided power readings from \textit{pm\_counters}, which is also used by Slurm. 
While Slurm only reports node-level power measurements, PMT can collect power measurement data for \textit{GPU}, \textit{CPU}, and \textit{memory}, and calculates the amount of energy consumed.
It is important to note that the energy consumption of the auxiliary parts of a node can also be calculated by subtracting the GPU, CPU, and memory measurements from the node-level energy measurement.

\textbf{Application.} SPH-EXA~\cite{cavelan2020sphexa} is a simulation framework that leverages state-of-the-art SPH method implementation.
It is written in modern C++, with minimal software dependencies and can execute on CPUs and GPUs at very large scales~\cite{keller2023cornerstone}, making it well-suited for extreme-scale astrophysical and cosmological simulations.

SPH-EXA uses MPI+X for parallelization, where X is either CUDA, HIP if the system offers CPUs and GPUs, and/or OpenMP if the system offers only CPUs.
SPH-EXA moves all the simulation data to the GPU memory at the beginning of the simulation and runs entirely on GPUs, leaving CPUs available for handling auxiliary tasks, such as performance or energy profiling. 
This way, the performance of the instrumented code is unaffected by performance or energy profiling.

\textbf{Measurement of application energy consumption.} 
SPH-EXA provides hooks within the code that can be used for low-overhead profiling, enabling third party tools to be integrated into the framework for performance and energy consumption analysis. 
The hooks are normally used to measure the timings for each function within the framework, which consists of functions that perform domain decomposition, halo exchanges, and physics computational kernels.
We used these hooks to instrument SPH-EXA function calls with PMT measurements to collect energy consumption from the start of each function call until its completion.
Energy consumption is measured per each MPI rank throughout the simulation, gathered at the end of the execution, and stored into a file for post-hoc analysis, to avoid perturbing the actual simulation.

The general rule-of-thumb in GPU-centric applications is to use one MPI rank to drive one GPU. 
However, this assignment does not favor energy measurements. 
For example, HPE/Cray \textit{pm\_counters} measure power consumption and report it in a per GPU card file.
On systems where each GPU card has two GCDs (GPU Complex Dies), such as LUMI-G (Table~\ref{table:systems}), an MPI rank only drives a GPU half-card, while power consumption is measured and reported for the entire card which corresponds to two MPI ranks.
This is not an issue on systems where each GPU card has one GCD, such as the CSCS-A100 system (Table~\ref{table:systems}). 
In both cases, \textit{pm\_counters} reports 4-or-8 power consumption measurements on CPUs, one for each MPI rank executing on one node.
The CPU measurements for both systems take into account the entire CPU, and all MPI ranks on the same node report the same energy measurement, hence only one measurement needs to be used for the calculations.
To overcome these discrepancies, our analysis scripts take the system's hardware configuration and MPI rank-to-GPU assignment into consideration.
However, two GCDs on one GPU card still creates certain measurement inaccuracies, as discussed in Section~\ref{subs:validation}.

With PMT integrated into SPH-EXA, we executed astrophysical simulations and collected the time and energy measurements throughout the runs.
The information gathered is exploited to draw insights from the energy consumption of the simulation at device-level and code functional-level.
We show that functional-level energy consumption can be used to designate which part of the simulation code can benefit from GPU frequency down-scaling for reducing the energy consumption.

\begin{table}[hb!]
    \centering
    \caption{Simulation and computing system parameters.}
    \label{table:systems}
    \resizebox{\columnwidth}{!}{%
    \begin{tabular}{|l|l|l|}
    \hline
    \textbf{Simulation} & \textbf{Parameters}            & \textbf{Info}                                    \\ \hline
Subsonic Turbulence & -n 0.6|1.2|2.4|4.9|7.4|9.2|14.7 Billion particles -s 100     & 150 million particles per GPU | 100 time-steps \\ \hline
Evrard Collapse     & -n 0.6|1.2|2.4|3.2|4.8|7.7 Billion particles -s 100  & 80 million particles per GPU | 100 time-steps          \\ \hline
    \textbf{System} & \textbf{Hardware of each Node} & \textbf{GPU Frequencies} \\ \hline
    LUMI-G & \begin{tabular}[c]{@{}l@{}}1x 64 cores AMD EPYC 7A53 CPU with 512 GB Memory\\ 8x AMD Mi250X GPUs half cards with 64 GB Memory\end{tabular} & \begin{tabular}[c]{@{}l@{}}AMD GPU compute frequency: 1700 MHz\\ AMD GPU memory frequency: 1600 MHz\end{tabular} \\ \hline
    CSCS-A100 & \begin{tabular}[c]{@{}l@{}}1x 64 cores AMD EPYC 7113 with\\ 4x NVIDIA A100-SXM4 with 80 GB Memory\end{tabular} & \begin{tabular}[c]{@{}l@{}}NVIDIA GPU compute frequency: 1410 MHz\\ NVIDIA GPU memory frequency: 1593 MHz\end{tabular} \\ \hline
    miniHPC & \begin{tabular}[c]{@{}l@{}}2 x 28 Core Intel Xeon Gold 6258R CPU with 1.5 TB Memory\\ 2 x NVidia A100-PCIE with 40GB Memory\end{tabular} & \begin{tabular}[c]{@{}l@{}}NVIDIA GPU compute frequency: 1410 MHz\\ NVIDIA GPU memory frequency: 1593 MHz\end{tabular} \\ \hline
    \end{tabular}%
    }
\end{table}

\section{Results}
\label{sec:results}

We conducted experiments with the SPH-EXA simulation framework and executed Subsonic Turbulence and Evrard Collapse simulations with 80 and 150 million particles per GPU, respectively, and evaluated the value added by PMT instrumentation, across three systems with GPUs as shown in Table~\ref{table:systems}.

We first validate the energy measurements from PMT-instrumented SPH-EXA against Slurm provided measurements on LUMI-G and CSCS-A100 systems.
We then show how the measurements with PMT enable important information on energy consumption per-device on the system as well as per-function during the simulation.
Furthermore, we use the miniHPC system for exploiting the PMT provided information to evaluate the effect of GPU frequency down-scaling on energy consumption.

\subsection{Validation of PMT Energy Measurements}
\label{subs:validation}

\begin{figure}[h]
    \centering
    \includegraphics[width=0.8\columnwidth]{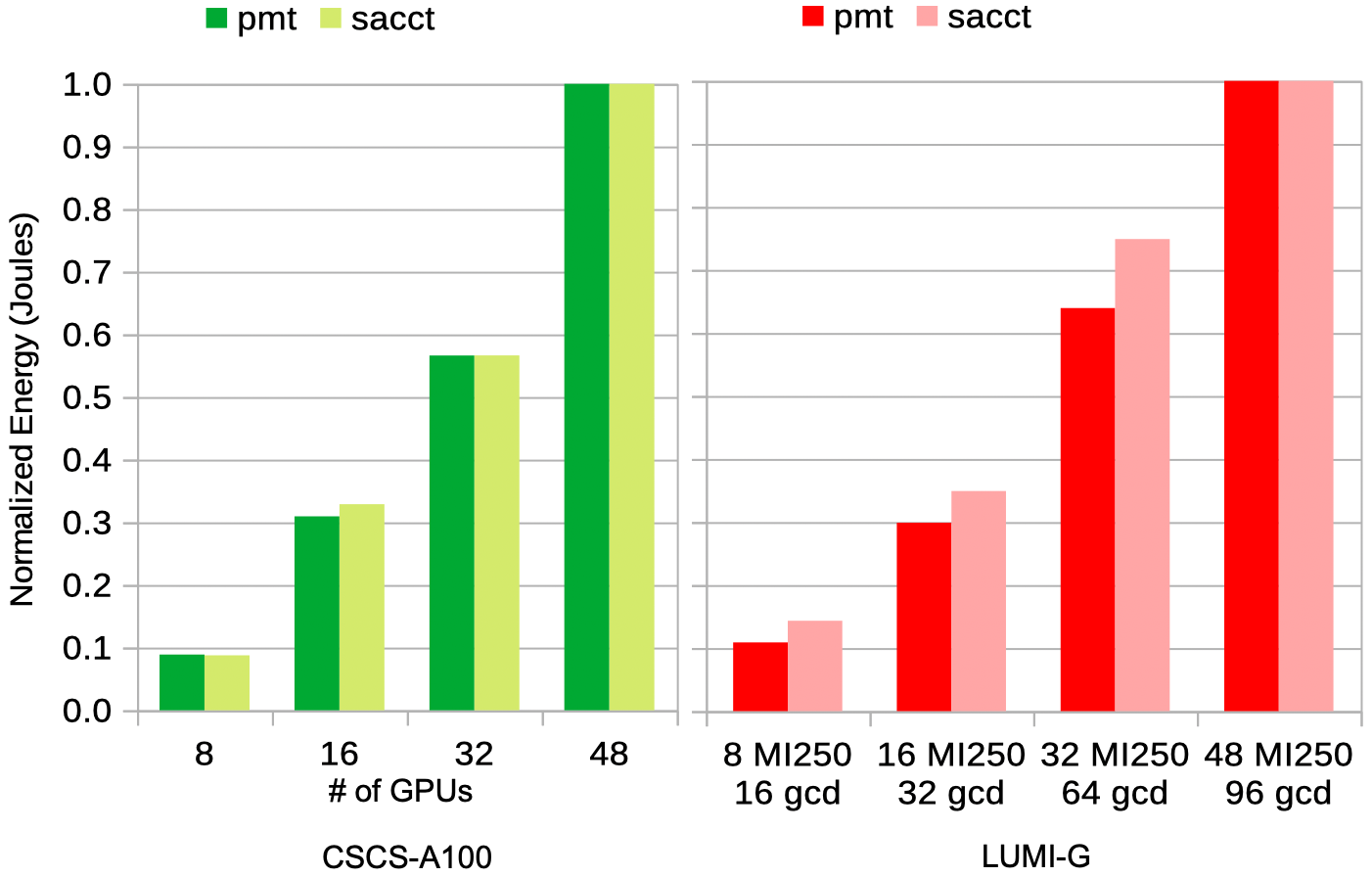}
    \caption{Comparison between PMT measured energy and Slurm reported energy.}
    \label{fig:pmtVsSlurm}
\end{figure}

Integration of instrumentation-based power measurement tools, such as PMT, does not guarantee the validity of the measurements. 
Since users typically only have access to Slurm measurements, we validated our PMT instrumented simulations to this baseline.
To this end, we validate the PMT measured energy with Slurm provided energy by running the Subsonic Turbulence experiments with energy measurement enabled, using 8-to-48 GPU cards with 1 GPU per MPI rank as shown in Figure~\ref{fig:pmtVsSlurm}.
The first observation is that the PMT energy consumption measurements are more accurate on the CSCS-A100 system, while energy consumption is underestimated by PMT on LUMI-G.

The difference between PMT and SLURM measurements is due to the timing of the energy measurement. 
Slurm starts measuring energy as soon as the job is submitted, while PMT starts the measurement when the time-stepping loop begins in SPH-EXA. 
This means that PMT does not measure the job and application setup phases, such as job launching or allocating the required data structures for the simulation. 
This difference in measurements does not pose a problem for the actionable insights users can take, as users have limited control over the job setup time and reducing the energy consumption of application initialization has a limited impact on the total amount of energy consumed.
This is because the computational units that consume the most energy, namely GPUs, are idle during job and application setup phases.

\begin{figure}[h]
    \centering
    \includegraphics[width=0.75\columnwidth]{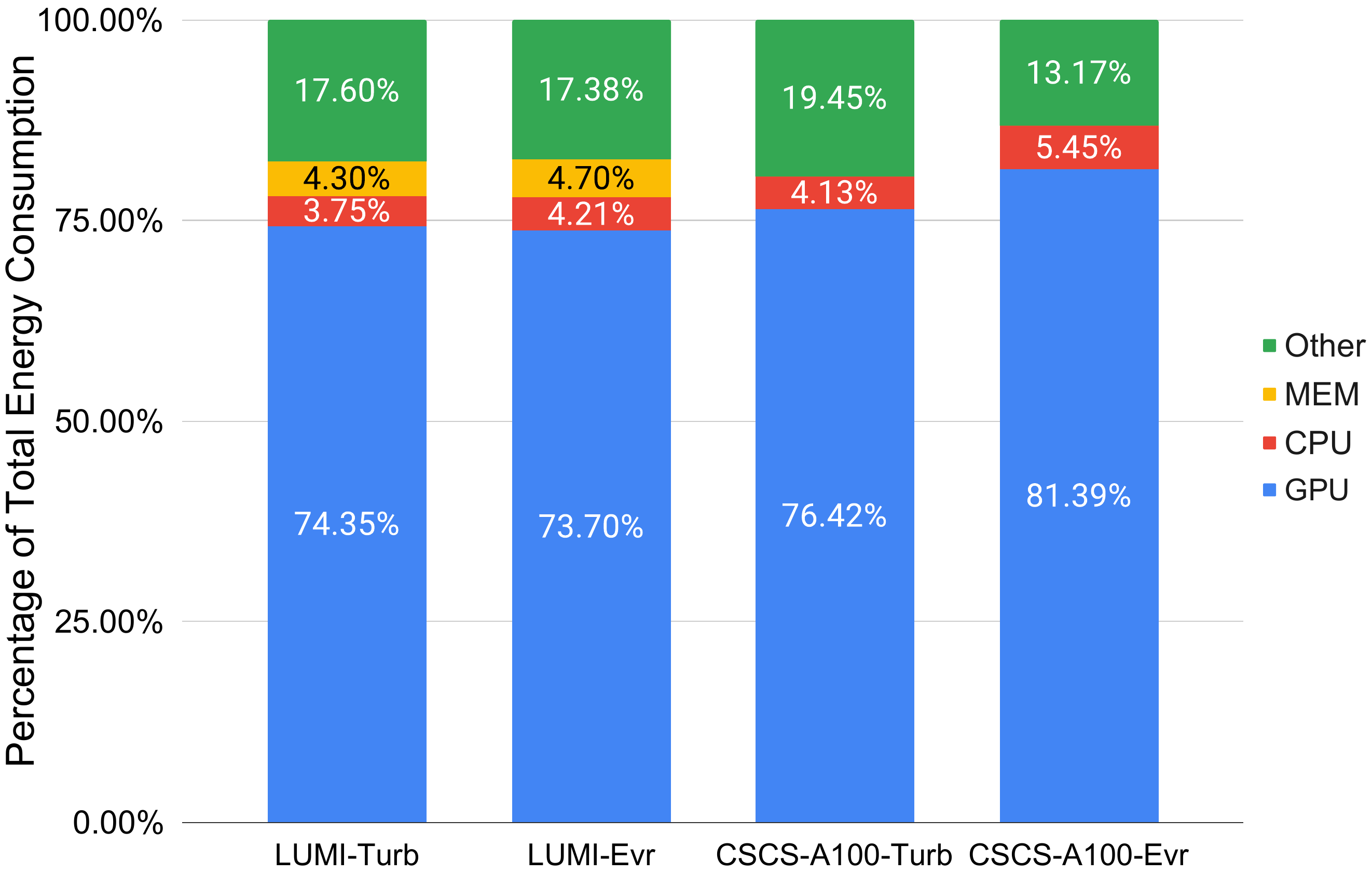}
    \caption{Device breakdown of consumed energy of Subsonic Turbulence and Evrard Collapse simulations on two systems.}
    \label{fig:dev_breakdown}
\end{figure}
Figure~\ref{fig:dev_breakdown} shows the percentage of energy consumed by each device on the two test systems for the Subsonic Turbulence and Evrard Collapse simulations.
The LUMI-G system shows separate measurements for GPU, CPU, and memory while CSCS-A100 system does not provide separate measurements for the  memory, hence the \textit{Other} reported for CSCS-A100 also includes the energy consumed by the memory.
The amount of total energy consumed in mega-Joules~(MJ) is 24.4, 15.2, 12.5, and 10.7 for LUMI-Turb, LUMI-Evr, CSCS-A100-Turb, and CSCS-A100-Evr, respectively.

The energy consumption breakdown by device shows that the GPU consumes the most energy, $74.3\%$ on LUMI-G and $76.4\%$ on CSCS-A100 system.
This information already provides insight about where the greatest potential lies for reducing energy consumption.
Since SPH-EXA computations are predominantly executed by the GPU, the optimizations focusing on energy efficiency need to be applied to the computational kernels.

The energy consumed by auxiliary parts of the node, named as \textit{other}, is a calculated value as explained in Section~\ref{sec:methodology}.
Even though it is the second-most energy consuming part of the simulation, we do not have additional information to insightfully analyze it.
For example, it would be important to know if the energy consumed is attributed to the network interface card, so that the communication operations become a target for future optimizations.

\begin{figure}[h]
    \centering
    \includegraphics[width=\columnwidth]{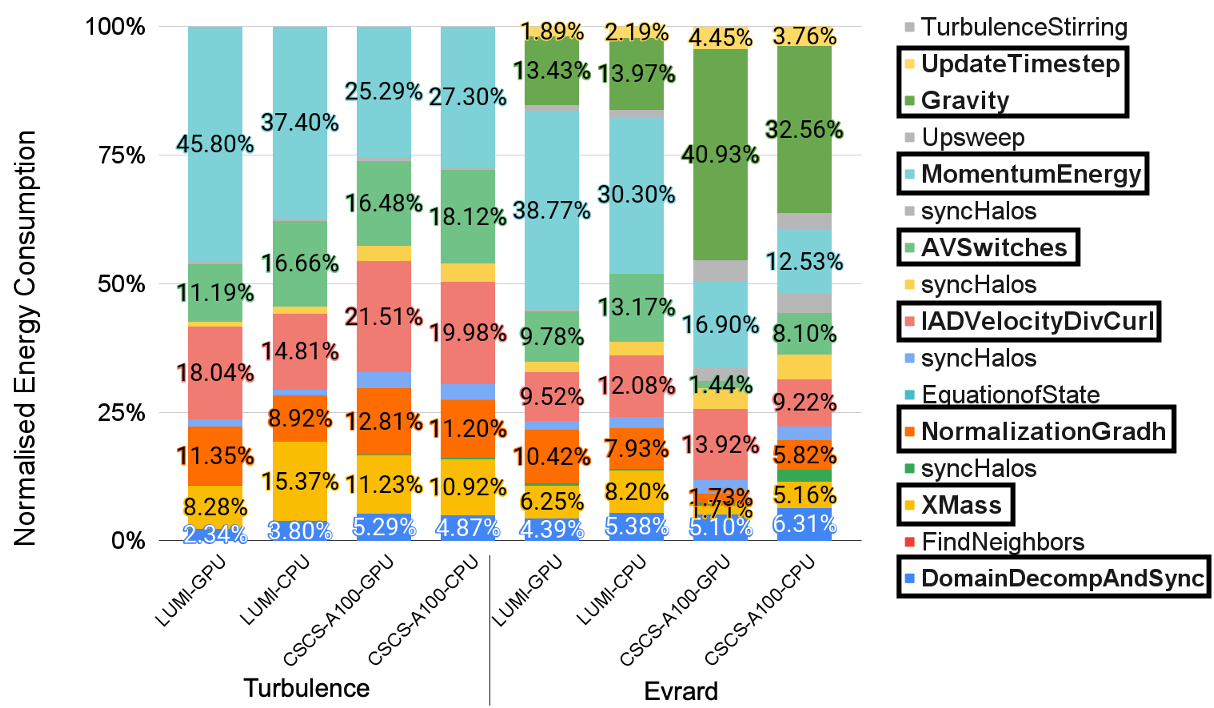}
    \caption{Breakdown of energy consumption by SPH-EXA functions for the Subsonic Turbulence and Evrard Collapse simulations executing on two systems.}
    \label{fig:devAndFunc}
\end{figure}

The instrumented code also enables energy measurements at code functional-level. 
Figure~\ref{fig:devAndFunc} shows all SPH-EXA functions called in the time-stepping loop of the Subsonic Turbulence and Evrard Collapse simulations. 
The functions that consume the most energy overall are enclosed in a box in the legend. 
The reason these functions also consume the most energy on the CPU as well as on the GPU, even though the CPU does not execute any work is due to the fact that CPU still consumes energy which is proportional to the execution time of each function.

The energy consumption per function normalized to the total energy consumed per device varies greatly depending on the system. 
For example, for the Turbulence simulation on the CSCS-A100 system, the function \textit{MomentumEnergy} only consumes $25.29\%$ of the total energy consumed by the GPU (3.1 MJ) while on LUMI-G it consumes $45.80\%$ (11.2 MJ).
This is a clear indication that \textit{MomentumEnergy} function can further be optimized for AMD GPUs.

\subsection{The Effect of GPU Frequency Changes}
\label{subs:freq}

The information made available by the integration of PMT into \mbox{SPH-EXA} shows where the most energy is consumed both device-wise and function-wise which is paramount for the energy efficiency optimization of simulations. 
Efficiency metrics such as energy-delay product, calculated by multiplying the total amount of energy with the execution time, can be used to quantify the impact of applied optimizations.

Previous studies~\cite{portegies2020ecological} show the effect of the choice of programming languages and compute device frequencies for running the simulations in an astrophysical context.
Zwart~et.~al.~\cite{portegies2020ecological}, show that GPUs are the most energy efficient for large scale simulations and they evaluate the effect of CPU frequency changes on the energy-to-solution. 
Since SPH-EXA already supports GPU execution, we employed frequency down-scaling as a way to reduce the GPU power consumption as other studies~\cite{ge2013effects} show the benefits of reducing the energy consumption through reduced power consumption.

\begin{figure}[h]
    \centering
    \includegraphics[width=0.9\columnwidth]{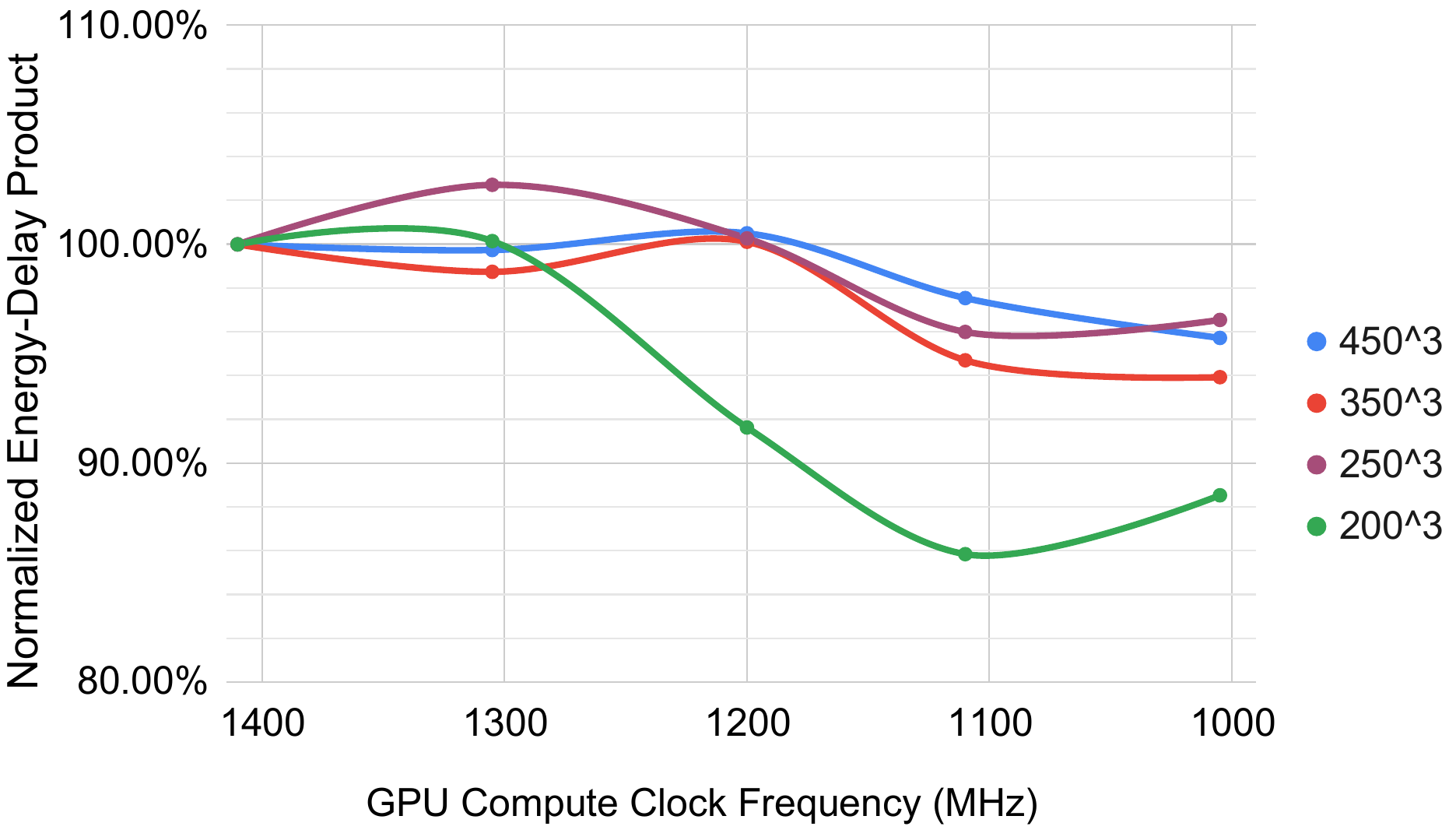}
    \caption{The effect of down-scaling frequency on the energy-delay product of Subsonic Turbulence simulation with different GPU particle counts.}
    \label{fig:a100edp}
\end{figure}

As the systems we used to gather the energy measurements in larger simulations do not allow user control over the GPU frequencies, we used the GPU node of \textit{miniHPC} to conduct experiments with different GPU compute frequencies. 
Since the GPU memory of miniHPC is smaller than the GPU memory on the other two test systems, we were forced to execute smaller simulations, starting at 91 million particles per GPU.

Figure~\ref{fig:a100edp} the energy-delay product (EDP) of SPH-EXA turbulence simulations normalized to the baseline, when the GPU compute frequency is set to 1'410 MHz.
We compare the EDP of SPH-EXA simulations when varying the GPU compute frequency between 1'410 MHz and 1'005 MHz. 
As frequencies are reduced, the time-to-solution of the simulation increases, but the power consumption reduction is so significant that the EDP decreases.

\begin{figure}[h]
    \centering
    \includegraphics[width=\columnwidth]{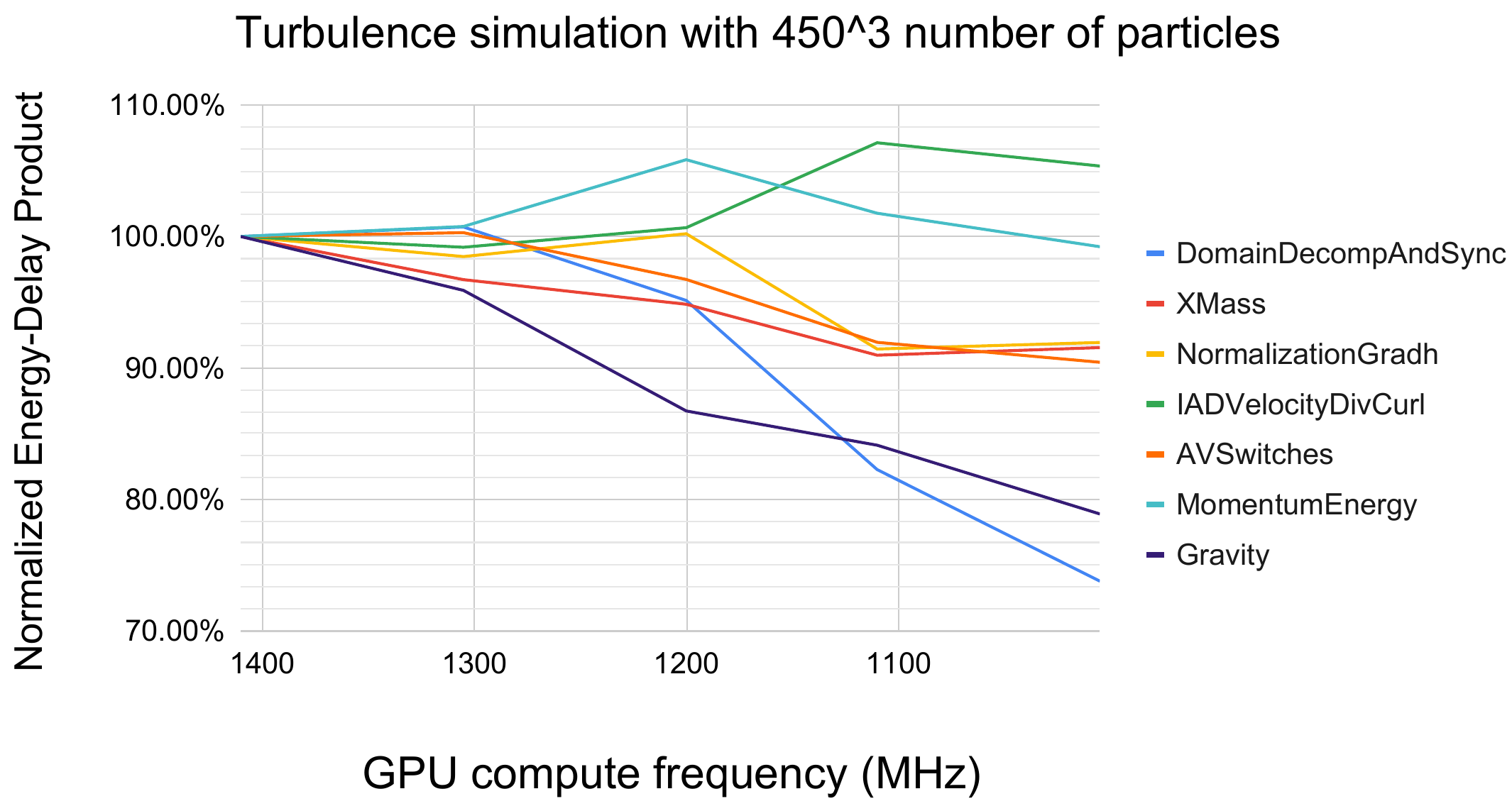}
    \caption{Energy-Delay product of the most time consuming functions of Subsonic Turbulence simulation running on different GPU compute frequencies.}
    \label{fig:edpFreq}
\end{figure}

We also evaluate the effect of the change in the simulation problem size on the EDP, by varying the simulated particles per GPU between 450\textsuperscript{3} (91 million) particles down to 200\textsuperscript{3} (8 million) particles.
We observe that the EDP drops significantly when the GPUs are not fully utilized as evidenced by the 200\textsuperscript{3} particle case in green.

Down-scaling the GPU frequency during the entire simulation to save energy is not the best solution as it increases the overall time-to-solution.
However, dynamic approaches can be employed for identifying Pareto-optimal solutions that trade-off high performance and low power consumption.

Figure~\ref{fig:edpFreq} shows the EDP of the most time consuming functions in SPH-EXA simulations with 450\textsuperscript{3} particles per GPU.
The most compute-bound functions, i.e., \textit{MomentumEnergy} and \textit{IADVelocityDivCurl}, do not benefit from reduced GPU compute frequency while the less compute-bound function \textit{DomainDecompAndSync} sees a $27\%$ reduction in its EDP.
The remaining functions also benefit by up to $20\%$ reduction in their respective EDP.

\section{Conclusion}
\label{sec:conclusion}

This paper presents the integration of an open-source power measurement toolkit, PMT, into SPH-EXA, an open-source astrophysics and cosmology simulation framework. 
We investigated the value this integration brings on systems with multiple compute devices, showing how much energy each device consumes, as well as the energy consumption of each simulation function.
The information made available can be used to identify parts of the simulation code that can be improved for performance and energy efficiency.

We first validated our approach using Subsonic Turbulence simulations, and compared the energy measurements from PMT with Slurm.
We also measured the energy consumption of Subsonic Turbulence and Evrard Collapse simulations on a per-function level in a system where the GPU frequency settings can be changed.
The results of our experiments show that some parts of the simulation can benefit from reduced GPU compute frequencies, making the code more energy-efficient.

While GPUs employ dynamic frequency scaling, this scaling is not as fine-grained as for CPUs.
Additionally, users may not have control over GPU DVFS, which hinders the potential energy-reducing optimizations.
Our results indicate that the integration of PMT into SPH-EXA provides valuable information for understanding and improving the energy efficiency of simulations.

Future work includes the utilization of the gathered data per-function and employing variety of dynamic approaches from the literature that trade-off high performance and energy consumption, bringing down the total energy consumed by the simulations.

\section*{Acknowledgements}
The work in this paper is supported by the Swiss Platform for Advanced Scientific Computing (PASC) project SPH-EXA2 (2021-2024) and as part of SKACH consortium through funding with The State Secretariat for Education, Research and Innovation (SERI) of Switzerland.

\bibliographystyle{ACM-Reference-Format}
\bibliography{references}

\end{document}